\documentclass[twocolumn,aps,showpacs,groupedaddress]{revtex4}
\usepackage{amssymb}
\usepackage{graphicx}
\usepackage{epsfig}
\usepackage{subfigure}

\begin{document}
\title{Nonlinear effect  on the transmission of light in a cavity array}
\author{H. D. Liu, W. Wang, and X. X. Yi\footnote{yixx@dlut.edu.cn}}
\affiliation{School of Physics and Optoelectronic Technology,
              Dalian University of Technology, Dalian 116024, China}

\date{\today}

\begin{abstract}
Taking nonlinear effect  into account, we study theoretically the
transmission properties of photons in a one-dimensional coupled
cavities, the cavity located at the center of the cavity array is
coupled to a two-level system. By the traditional scattering theory,
we calculate the transmission rate of photons along the cavities,
and discuss the effect of nonlinearity on the  photon transport. The
results show that the controllable two-level system can act as a
quantum switch in the coherent transport of photons. The dynamics of
such a system is also studied by numerical simulations,  the effect
of the atom-field detuning and  nonlinearity on the dynamics is
shown and discussed.
\end{abstract}
 \maketitle

\section{Introduction}
Recent experimental progress in the fabrication of microcavity
arrays and the realization of the quantum regime in the coupling of
atomic-like structures to quantized electromagnetic modes inside the
cavities
\cite{armani03,aoki06,barclay06,akahane03,hennessy07,wallraff04,trupke07,colombe07}
open up the possibility of using them as quantum simulators of
many-body physics. The possibility to create such cavity arrays have
stimulated a number of theoretical investigations on the transport
physics\cite{fan05,sun08,xu00}, with particular emphasis on possible
analogies with mesoscopic phenomena in the electronic context. This
study is motivated by the preparation of new quantum states for
photons or combined photonic-atomic excitations, which offer
advantages over other systems for the realization of strongly
interacting many-body models in quantum optical systems. For
Josephson junction arrays and optical lattices, it is  difficult to
address individual sites due to the small separations between
neighboring sites, but this is easy for coupled cavity arrays due to
the large separations which are usually dozens of micrometers and
can therefore be accurately accessed by optical frequencies.

Coupled cavity arrays have several interesting potential
applications, including quantum information processing and
simulations of quantum strongly correlated many-body systems. By
using a real-space model Hamiltonian, it has been shown that the
transmission of a single photon can be switched on or off in
one-dimensional waveguide coupled with superconductivity quantum
bits\cite{fan05}. Indeed the transmission can be switched to any
predicted value with a tiny change in the magnetic field when the
 sharper Fano resonance peak is employed. For a single photon
inside a one-dimensional resonator waveguide with  a two-level
system, a general spectral structure in which the reflection and
transmission beyond the usual Breit-Wigner and Fano line shapes is
predicted\cite{sun08}. The atom in a half-waveguide can play an
intrinsic role of  semi-transparent mirror for the single photon,
leading the system to be flexible under control.

Nonlinearity may appear in quantum systems due to the many-body
effects and/or system-environment couplings. Transport and tunneling
properties of nonlinear system (for example, Bose-Einstein
condensates) are of considerable current interests, both
experimentally and theoretically. Quantum tunneling through a
barrier is a paradigm of quantum mechanics and usually takes place
on a nanoscopic scale, such as in two supperconductors separated by
a thin insulator\cite{likharev79} and two reservoirs  of superfluid
helium connected by nanoscopic apertures\cite{pereverzev97,
sukhatme01}. Recently, tunneling on a macroscopic scale ($\mu m$) in
two weakly linked Bose-Einstein condensates in a double-well
potential has been observed\cite{albiez05}. Similar to tunneling
oscillations in superconducting and superfluid Josephson junctions,
Josephson oscillations  are observed when the initial population
difference is chosen to be below a critical value. When the initial
population difference exceeds the critical value, an interesting
feature can be observed, i.e., tunneling oscillations are suppressed
due to the nonlinear condensate self-interactions. This phenomenon
is known as the macroscopic quantum self-trapping.

Here we explore a different regime of transport for  photons in a
cavity array coupled to a two-level system. Nonlinear effects
arising from  Kerr medium in cavities are taken into account. The
paper is organized as follows. In Sec.{\rm II}, we present a general
formulism for the atom-cavity array. By the traditional scattering
theory, the transmission rate is calculated and discussed in Sec.
{\rm III}. In Sec. {\rm IV}, we present a numerical simulation for
the dynamics of the coupled atom-cavity array system. An
experimentally realizable proposal to observe the prediction is
suggested in Sec.{\rm V}. Finally, we conclude our results in
Sec.{\rm VI}.

\section{Formulism}
\begin{figure}
\vskip 1 cm
\includegraphics*[width=0.8\columnwidth,
height=0.15\columnwidth]{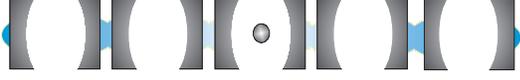} \caption{(color
online)Schematic illustration of the coupled cavity arrays, in which
the photons propagate. A two-level atom is located at the zeroth
cavity and coupled to the cavity field.  } \label{fig1}
\end{figure}

Consider a cavity array coupling to a two-level system.
 Photon hopping can occur between neighboring cavities
due to the overlap of the special profile of the cavity modes, see
figure \ref{fig1}.
Introducing the creation and annihilation operators of the cavity
modes, $a_j^{\dagger}$ and $a_j$, the Hamiltonian of the field in
the cavity can be written as,
\begin{equation}
H_{c}^l=\omega\sum_{j}a_{j}^{\dag}a_{j}-\xi\Sigma_{j}(a_{j}^{\dag}a_{j+1}+h.c.).
\label{lcavity}
\end{equation}
In Hamiltonian Eq.(\ref{lcavity}) that all the cavities have the
same resonant frequency and the same spacing overlap between  all
neighboring cavities have been assumed. In contrast to the previous
studies, we focus here on the influence of nonlinearity on the
transparent properties of  photons in the cavity array. The
nonlinearity comes from the optical Kerr effects, where the
electronic field is due to the light itself. Taking this
nonlinearity into account, we add an additional term
$g\sum_j(a^{\dagger}a)^2$ into the Hamiltonian, and the Hamiltonian
for the cavity array therefore becomes
\begin{equation}
H_{c}=H_c^l+ g\sum_{j}(a_{j}^{\dag}a_{j})^{2}.
\end{equation}
The location of the two-level atom in the cavity array is chosen to
be the origin of the coordinate axis, which is only coupled to the
zeroth cavity field in the cavity array.  We assume that the cavity
decay and atomic spontaneous emission are ignored. This
configuration is different from that in
Refs\cite{zhou07,hu07,hartmann06,rakhmanov08}, where several
approximations have been made to treat the problem. The coupling of
the two-level atom to zeroth cavity field is described by the
Janes-Cummings model(under the Rotating Wave Approximation),
\begin{equation}
H_{I}=\Omega|e\rangle\langle e|+J_0(a_{0}^{\dag}|g\rangle \langle
e|+|e\rangle\langle g|a_{0}),
\end{equation}
where $J_0$ denotes the coupling constant. To observe the effect of
this nonlinearity, we consider the situation where the field
intensities are very high, this yields
\begin{eqnarray}
i\hbar\dot{\alpha}_{j}&=&\omega\alpha_{j}-\xi(\alpha_{j+1}+\alpha_{j-1})
+2g|\alpha_{j}|^{2}\alpha_{j},\nonumber\\
i\hbar\dot{\alpha}_{0}&=&\omega\alpha_{0}-
\xi(\alpha_{-1}+\alpha_{1})+2g|\alpha_{0}|^{2}\alpha_{0}+
J_0\langle\sigma_{-}\rangle,\nonumber\\
i\hbar\langle\dot{\sigma}_{-}\rangle&=&\Omega\langle\sigma_{-}\rangle
-J_0\alpha_{0}\langle\sigma_{z}\rangle,\nonumber\\
i\hbar\langle\dot{\sigma}_{z}\rangle&=&-2J_0\alpha_{0}^{*}
\langle\sigma_{-}\rangle+2J_0\alpha_{0}\langle\sigma_{+}\rangle,\nonumber\\
i\hbar\langle\dot{\sigma}_{+}\rangle&=&-\Omega\langle\sigma_{+}\rangle
+J_0\alpha_{0}\langle\sigma_{z}\rangle.\label{mainec}
\end{eqnarray}
Here, to account for the classical and quantum fluctuations, each
operator was decomposed as a sum of its average value and a small
fluctuation, i.e., $a_j=\alpha_j+\delta a_j,$ and
$\sigma_{+,-,z}=\langle \sigma_{+,-,z}\rangle+\delta
\sigma_{+,-,z}.$ Substituting these quantities into the Heisenberg
equation, $i\hbar\partial A/\partial t=[A, H_c+H_I]$, $(A=a_j,
\sigma_+, \sigma_-,\sigma_z, j=0,1,2,...)$, we have
Eq.(\ref{mainec}) and the following  equation for the fluctuations,
\begin{eqnarray}
i\hbar \frac{\partial}{\partial t}\delta a_{j}&=&\omega \delta
a_{j}-\xi(\delta a_{j+1}+\delta a_{j-1})\nonumber\\
&+&4g|\alpha_j|^2\delta a_{j}+2g\alpha_j^2\delta a_j^{\dagger},
\quad(j\neq 0),\nonumber\\
i\hbar \frac{\partial}{\partial t}\delta a_{0}&=&\omega
\delta a_{0}-\xi(\delta a_{1}+\delta a_{-1})+J_0\delta\sigma_-\nonumber\\
&+&4g|\alpha_0|^2\delta a_{0}+2g\alpha_0^2\delta a_0^{\dagger},\nonumber\\
i\hbar \frac{\partial}{\partial
t}\delta\sigma_{-}&=&\Omega\delta\sigma_{-}-J_0\alpha_0
\delta\sigma_{z}
-J_0 \delta a_0\langle\sigma_{z}\rangle,\nonumber\\
i\hbar\frac{\partial}{\partial t}{\delta\sigma}_{z}&=&-2J_0(\delta
a_{0}^{\dag}\langle \sigma_{-}\rangle -\alpha_{0}\delta \sigma_{+}+
\alpha_{0}^{*}\delta\sigma_{-}
-\delta a_{0}\langle\sigma_{+}\rangle),\nonumber\\
i\hbar \frac{\partial}{\partial
t}\delta\sigma_{+}&=&-\Omega\delta\sigma_{+}+J_0\alpha_{0}^*\delta\sigma_{z}
+J_0\delta a_{0}^{\dagger}\langle\sigma_{z}\rangle.\label{fluc}
\end{eqnarray}
In deriving Eq.(\ref{fluc}) we have eliminated the average value
contribution and linearized the fluctuations. The dynamical
stability of the system can be obtained by Eq.(\ref{fluc}), this
method adopts the linear stability analysis that has wide
applications in various nonlinear systems. Eq.(\ref{fluc}) can be
written as $i\hbar \frac{\partial \vec{V}}{\partial t} =H_{eff}
\vec{V}$ with $\vec{V}=(\delta a_{-N},\delta a_{-N}^{\dagger},
\delta a_{-N+1},\delta a_{-N+1}^{\dagger},$ $...,\delta a_0, \delta
a_0^{\dagger},...,$ $\delta a_{N-1},\delta a_{N-1}^{\dagger}, \delta
a_{N},$ $\delta a_{N}^{\dagger},\delta \sigma_-, \delta \sigma_z,$
$\delta \sigma_+)^T$, and $H_{eff}$ can be derived from
Eq.(\ref{fluc}). We shall not present the details of  analysis here,
which will be discussed elsewhere.

\section{Transmission rate}

 Now we turn to the
shifted picture defined by the transformation, $\langle
\sigma_-\rangle \rightarrow \langle \tilde{\sigma_-}\rangle =\langle
\sigma_-\rangle e^{i\Omega_k t},$ $\Omega_k=-2\xi\cos
k+\omega+2g|\alpha_j|^2,$ by setting $\partial \langle
\tilde{\sigma_-}\rangle/\partial t=0,$ and solve for $\langle
\tilde{\sigma_-}\rangle$ in terms of $\langle {\sigma_z}\rangle,$ we
arrive at
\begin{eqnarray}
&&\Omega_{k}\alpha_{0}=\omega\alpha_{0}-\xi(\alpha_{-1}
+\alpha_{1})+2g|\alpha_{0}|^{2}\alpha_{0}+\alpha_0J_0f(\Delta,J_0),\nonumber\\
&&\Omega_{k}\alpha_{j}=\omega\alpha_{j}-\xi(\alpha_{j+1}
+\alpha_{j-1})+2g|\alpha_{j}|^{2}\alpha_{j},\nonumber\\
\label{zeroe}
\end{eqnarray}
where $f(\Delta,J_0)=\frac{J_0\langle
{\sigma_z}\rangle}{\Omega-\Omega_k},$ and $\Delta=\Omega-\Omega_k.$
Equations (\ref{zeroe}) represent a coupled set of stationary states
of the total system for the fields in the high-intensity limit. The
scattering equation for $j\neq 0$ has the solution,
\begin{eqnarray}
&&\alpha_{j}^{L}(k)=e^{ikj}+re^{-ikj},\quad j<0 \nonumber\\
&&\alpha_{j}^{R}(k)=se^{ikj},\qquad\qquad j>0,
\end{eqnarray}
where $s$ and $r$ denote the transmission and reflection amplitude,
respectively. By solving the scattering equation with the continuous
condition $\alpha_0^R(k)=\alpha_0^L(k)$, we obtain the transmission
amplitude $s$ satisfying,
\begin{eqnarray}
& \ &2g|s|^{2}\cdot s+(2\xi\cos
k+\frac{J^{2}}{\Omega-\Omega_{k}}-2\xi
e^{ik})s\nonumber\\
&\ &+2i\xi\sin k=0, \label{seq1}
\end{eqnarray}
where $J=J_0\sqrt{\langle \tilde{\sigma_z}\rangle}.$ Set $s=x+iy$
with real numbers $x$ and $y$, Eq.(\ref{seq1}) follows,
\begin{eqnarray}
&&2g(x^{2}+y^{2})y+(\frac{J^{2}}{\Omega_{k}-\Omega})y-2x\xi\sin
k+2\xi\sin k=0,\nonumber\\
&&2g(x^{2}+y^{2})x+(\frac{J^{2}}{\Omega_{k}-\Omega})x+2y\xi\sin k=0,
\label{seq2}
\end{eqnarray}
Eq.(\ref{seq2}) are coupled cubic equations giving  three roots to
the transmission amplitude $s$. Considering the restrictions on the
amplitude $s$ (i.e., $x,y$ are real number and $x^2+y^2\leq 1$), we
have performed extensive numerical calculations for the transmission
amplitude $s$, selected results are presented in figure 2-5.

\begin{figure}
\includegraphics*[width=0.9\columnwidth,
height=0.6\columnwidth]{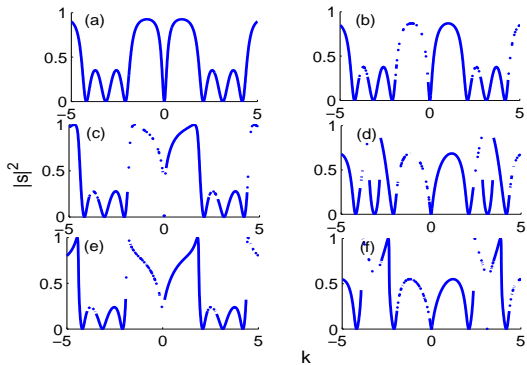} \caption{(color online) The
transmission coefficient $|s|^2$ as a function of the momentum $k$
of the incident photons. (a)-(d) correspond to different nonlinear
coupling constant $g=0,0.1,-0.5,0.5,-1,$ and $1,$ respectively. The
other parameters chosen are $J=1, \xi=1, \omega=2,$ and  $\Omega=3.$
} \label{fig2}
\end{figure}

Figure \ref{fig2} shows the transmission coefficient $|s|^2$ as a
function of the momentum $k$ of the incident photons. Figure
\ref{fig2}-(a) illustrates the transmission coefficient $|s|^2$
without nonlinear interactions, while figure \ref{fig2}-(b),(c) and
(d) show the effect of nonlinear coupling on the  transmission
coefficient $|s|^2$ with coupling constant $g= 0.1, -0.5, 0.5,-1$
and $1,$ respectively.  We can find from \ref{fig2}-(b),(c) and (d)
that the lines are not continuous, indicating no solutions to
Eq.(\ref{seq2}) can be found at the blank points. This is different
from the case without nonlinear couplings (namely, $g=0$).  Figure
\ref{fig2} also shows that the transmission coefficients not only
depend on the strength of the nonlinear coupling, but also on the
feature of the nonlinear interaction, i.e., the transmission
coefficients behave different for attractive and repulsive nonlinear
interactions (see, (c) and (d), (e) and (f)).

\begin{figure}
\includegraphics*[width=0.8\columnwidth,
height=0.6\columnwidth]{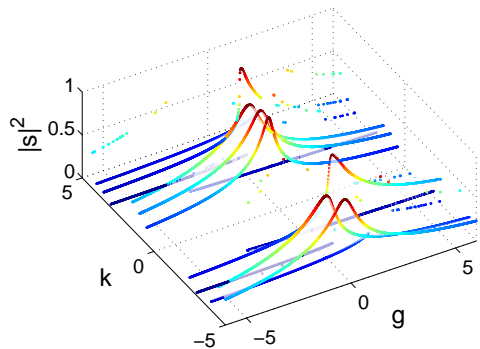} \caption{(color online) The
transmission coefficient $|s|^2$ as a function of the momentum $k$
and the nonlinear coupling constant $g.$  $J=1, \xi=1, \omega=2,$
and $\Omega=3$ were chosen for this plot. } \label{fig3}
\end{figure}
For large nonlinear coupling constants, the transmission coefficient
$|s|^2$ is zero, as figure \ref{fig3} shows. This means that the
photons can not transmitted along the cavity array.
\begin{figure}
\includegraphics*[width=1\columnwidth,
height=0.8\columnwidth]{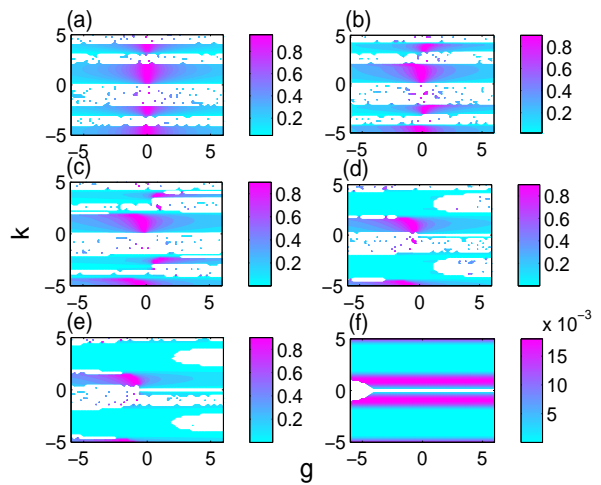} \caption{(color online)This
contour plot shows the transmission coefficient as a function of $k$
and $g$ with different cavity-atom coupling constant $J$. (a) $J=0,$
(b) $J=0.5$, (c) $J=1,$  (d) $J=1.5,$ (e) $J=2,$ and (f) $J=5.$ The
other parameters chosen are  $\xi=1, \omega=2,$ and $\Omega=3$. Note
that the blank regions in the figure indicate no solutions have
found for $|s|^2$.} \label{fig4}
\end{figure}
For $g=0$ and $J=0,$ the transmission coefficient $|s|^2$ is 1,
however this is not the case for nonzero $g$ and $J=0,$ $|s|^2$
would depend on $k$. For small $g$, the nonlinear effects block the
photon transport in most cases, however, for some special $k$, the
nonlinearity favors the photon transmission. For large $g$, the
nonlinear effect always decreases the transmission rate independent
of what $k$ takes.   The coupling of the atom to the photons can
shift the peaks of the transmission spectrum, this is shown  in
figure \ref{fig4}.
\begin{figure}
\includegraphics*[width=1\columnwidth,
height=0.8\columnwidth]{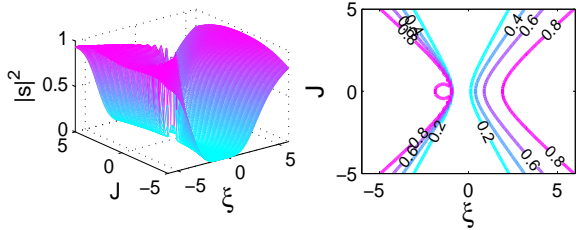} \vskip -3cm \caption{(color
online)Transmission coefficient as a function of $J$ and $\xi$ with
$k=1,g=1,\omega=2,$ and $\Omega=3.$} \label{fig5}
\end{figure}
For fixed $k$ and $g$, the dependence of the transmission
coefficient $|s|^2$ on $J$ and $\xi$ is shown in figure \ref{fig5}.
Figure \ref{fig4} shows that the peak of the transmission spectrum
is symmetric about $g=0$ when $J=0.$ As the coupling constat $J$
increases, the peaks are shifted and lose the symmetry about $g$,
i.e., the stronger the atom-field coupling, the bigger the peak
shift. Two observations can be found from figure \ref{fig5}. (1) For
a fixed $J$, the transmission coefficient $|s|^2$ increases with
$\xi$. This can be easily understood as that large $\xi$ would
increase the photon transmitting, resulting in larger $|s|^2.$ (2)
For a fixed $\xi$, the dependence of the transmission coefficient
$|s|^2$ on $J$ behaves different for positive and negative $\xi$.
For $\xi >0$, $|s|^2$ increases as $|J|$ decreases, whereas for
$\xi<0$, $|s|^2$ becomes larger with $|J|$ decreases.

\section{dynamics}
\begin{figure}
\includegraphics*[width=0.8\columnwidth,
height=0.6\columnwidth]{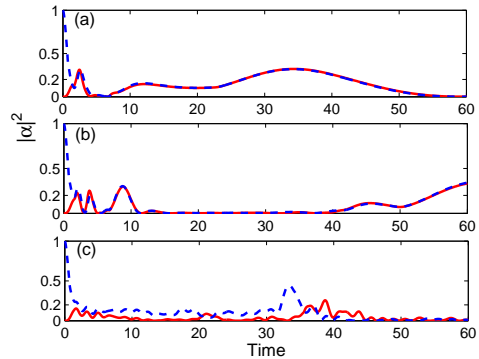}  \caption{(color online) Photon
number in $-1$th cavity (blue and dashed line) and $+1$th cavity
(red and solid line) as a function of time. The parameters chosen
are $\omega_r=2,$  $J_r=15,$ $M=15,$ $\xi_r=1,$ $g_r=2.$ (a)
$\Omega_r=2,$ (b)$\Omega_r=3,$  and (c) $\Omega_r=5.$} \label{fig7}
\end{figure}

\begin{figure}
\includegraphics*[width=0.8\columnwidth,
height=0.6\columnwidth]{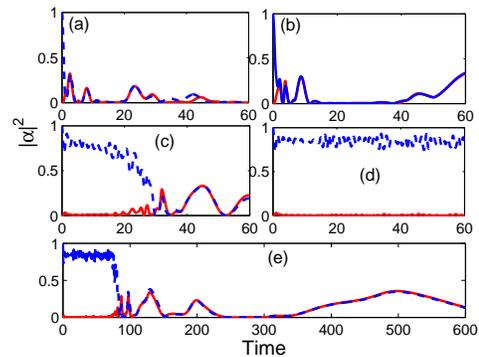}  \caption{(color online) The
same as Fig.\ref{fig7}, but with different parameters. $\omega_r=2,$
$\Omega_r=3.$  (a) $g_r=0.5,$ (b) $g_r=2,$ (c) $g_r=2.9,$ (d)
$g_r=3.$ (e) is a long time plot for (d).} \label{fig8}
\end{figure}

To gain insight of the nonlinear effect on the transport of photons
along the cavity array, we numerically simulate the dynamics of the
system, and present the numerical results in this section.  For this
purpose, we first rescall  the Hamiltonian $H=H_c+H_I$ by the total
number of photons in these cavities. Notice that for a very large
number of photons, $M=\langle \hat{M}\rangle \gg 1$ with
$\hat{M}=\sum_i a_i^{\dagger}a_i,$ the total number of photons is
(approximately) a constant of motion for the system. So, we may
simulate the dynamics  by using a Hamiltonian obtained by replacing
$a_i^{\dagger}, a_i$ with  $a_i^{\dagger}/\sqrt{M}, a_i/\sqrt{M},$
respectively,   this leads to a set of coupled equations which have
the same form as in Eq.(\ref{mainec}), but with rescaled constants,
$\omega_r=\omega, $ $\xi_r=\xi,$ $g_r=M \cdot g,$ $\Omega_r=\Omega,$
and $J_r=J/\sqrt{M}.$ We plot the photon number in the $-1$th (blue
and dashed line) and $+1$th (red and solid line) cavities as a
function of time in Fig.\ref{fig7} and Fig.\ref{fig8}. The initial
state of photons is all photons (total number $M=15$) in the $-1$th
cavity, while the atom initially is in its excited state
$|e\rangle.$ The $-1$th cavity is the first cavity left of the atom,
and the $+1$th one is the first on the right, the atom occupies the
$0$th cavity, as we described before. Fig.\ref{fig7} shows that for
resonantly atom-field coupling (Fig.\ref{fig7}-(a)), the photon in
the both cavities arrive in a synchronizely changed state for a very
short time. As the detuning $(\omega-\Omega)$ increases, it needs a
longer time to evolve into such a state (see
Fig.\ref{fig7}-(b),(c)). We also find from Fig.\ref{fig7} that for
resonantly atom-field coupling, it is easier for the photons to
transfer through the $0$th cavity than the case of non-resonant
coupling. This can be found by observing the first peak of the
red-solid curve, obviously, the first peak of the red-solid line in
Fig.\ref{fig7}-(a) is higher than that in (b) and (c).
Fig.\ref{fig8} shows the nonlinear effect on the dynamics of the
system. Again, the nonlinearity affects the time needed for the
cavity field to evolve into a synchronized state. The larger the
nonlinear coupling constant $g$ is, the longer the time to arrive at
such a state. Fig.\ref{fig8} also shows that the nonlinear effect
blocks the photon transfer along the cavity array with strong
nonlinearity, reminiscent of the self-trapping for Bose-Eintein
condensates in a double-well potential.

\section{Possible Experimental observation}
\begin{figure}
\includegraphics*[width=0.8\columnwidth,
height=0.2\columnwidth]{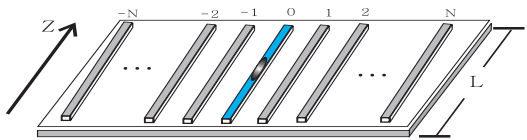}  \caption{(color online)A
schematic illustration of the setup. The relative distance between
the coupled waveguides equal to each other, resulting in the same
coupling constant among the neighboring waveguides. The two-level
atom was put in the zeroth waveguide in the center. } \label{fig6}
\end{figure}

In this section,  we propose an experimentally accessible quantum
system to demonstrate our theoretical predictions for the
transmission coefficient $|s|^2.$ Consider a system of $(2N+1)$
single-mode coupled nonlinear waveguides ( denoted as $-N, -N+1,
-N+2,...,-1,0,1,...,N-1,N,$ see figure \ref{fig6}). The
waveguides are identical in shape and have a constant
width\cite{lahini08} along the propagation direction $z$. The
distances between the waveguides equal to each other, as a
consequence, the coupling constants between the waveguides remain
unchanged along the propagation. $L$ denotes the waveguide length.
We assume that only couplings between neighboring waveguides are
practically nonzero in this configuration. The two-level atom is put
in the zeroth waveguide. The waveguides can be designed to have a
width of $3\mu m$ and length of $18 mm.$ The edge-to-edge distance
between waveguides is $5\mu m$, yielding a coupling constant of 2500
$m^{-1}$. The evolution of the modal amplitudes in those waveguides
can be described by Eq.(\ref{mainec}) in the intensive light limit.
The nonlinear couplings $g$ stem from the Kerr nonlinear coefficient
of the waveguides. These couplings are important only in the
nonlinear regime and can be neglected at low light power levels. We
would like to note that $\dot{A}$ in Eq.(\ref{fluc}) should be
understood as $\partial A/\partial z$ $(A=a_j, \sigma_+,
\sigma_-,\sigma_z, j=0,1,2,...)$ in this coupled waveguide system.
The waveguides used in this proposal can be fabricated on an
$AlGaAs$ substrate, using standard photolithography
techniques\cite{mandelik05}. Indeed by this configuration in three
coupled waveguides, the effect of nonlinearity on adiabatic
evolution of of light was observed\cite{lahini08}. Besides, these
features that result from the effect of nonlinearity can also
hopefully be observed in coupled superconducting transmission line
resonators.

\section{conclusion}

To sum up, taking nonlinear couplings of photons into account, we
have studied theoretically the transmission properties of intensive
light in a one-dimensional waveguide. The transmission coefficient
was calculated and discussed. These results show that the nonlinear
couplings sharply influence the transmission, and the atom-field
couplings shift the peaks in the transmission spectrum. The
dependence of the transmission coefficient on the couplings between
the neighboring sites is also calculated and analyzed. Besides, we
numerically simulate the dynamics of the system. The following
features are found. (1) Resonant atom-field coupling favors the
photon transport, the larger the detuning is, the longer the time
needed to arrive at a synchronized state. (2) Nonlinearity blocks
the transfer of photons among the cavity array, and postpone the
time to arrive at a synchronized state. (3) Both a off-resonant
atom-field coupling and a strong  nonlinearity make the photon
transport along the cavity array difficult with respect to the case
of resonant atom-field coupling and without (or small) nonlinearity.

This work was supported by   NSF of China under Grant  No. 10775023.\\

\end{document}